\documentclass[conference]{IEEEtran}
\IEEEoverridecommandlockouts
\usepackage{amsmath,amssymb,amsfonts}
\usepackage{algorithmic}
\usepackage{algorithm}
\usepackage{array}
\usepackage{textcomp}
\usepackage{stfloats}
\usepackage{url}
\usepackage{verbatim}
\usepackage{graphicx}
\usepackage{cite}
\usepackage[subtle,tracking=normal]{savetrees}
\usepackage{subcaption}
\usepackage{xcolor}
\usepackage{caption}
\usepackage{multirow}
\usepackage[left=0.68in,right=0.673in,top=0.7in,bottom=1.013in]{geometry}
\def\BibTeX{{\rm B\kern-.05em{\sc i\kern-.025em b}\kern-.08em
    T\kern-.1667em\lower.7ex\hbox{E}\kern-.125emX}}

\begin{document}
\title{A Simple Numerical Method for Non-Gaussian Signal Ensembles in Nonlinear Power Amplifiers}
\author{\IEEEauthorblockN{Cameron~M.~Pike\IEEEauthorrefmark{1}\IEEEauthorrefmark{2}, and
Animesh Yadav\IEEEauthorrefmark{2}\\
\IEEEauthorblockA{\IEEEauthorrefmark{1}GIRD Systems, Inc. Cincinnati, OH 45246, USA, and \IEEEauthorrefmark{2}Ohio University, Athens, OH  43147, USA\\
Email: cpike@girdsystems.com,
yadava@ohio.edu}}}


\maketitle

\begin{abstract}
Beam tracking in vehicular communication systems is inherently challenging due to high mobility and the use of narrow millimeter-wave (mmWave) beams. These challenges are further exacerbated by power amplifier (PA) nonlinearities, which introduce distortion-induced beam pattern deviations, array-gain loss, and non-Gaussian signal distortions. Motivated by the need for analytical tools capable of characterizing such effects, this paper extends Rice’s characteristic-function (ch. f.) method for the stochastic analysis of signals and noise in memoryless nonlinear systems. The proposed approach represents the nonlinearity using a Fourier series rather than a Fourier transform, transforming the evaluation of output correlation functions from computationally intensive double or triple improper integrals into tractable summations. The resulting framework preserves the generality of the original method, supporting one or more sinusoidal signals and noise processes that are not restricted to Gaussian distributions. A new fundamental ch. f.-based formulation is derived in terms of Fourier-series coefficients and a discrete parameterization of the generalized characteristic function. Numerical results are presented for a nonlinear GaN HEMT transconductance characteristic driven by a sinusoidal signal and Gaussian noise, demonstrating the applicability of the proposed method. The framework provides a computationally efficient tool for analyzing nonlinear RF front-end impairments and their impact on future wireless and vehicular communication systems.
\end{abstract}


\section{Introduction}
Nonlinearities are inherent in modern communication systems, particularly in 5G/6G networks that employ wideband signaling, high-order modulation, and energy-efficient hardware. At the transmitter, they primarily arise from radio frequency (RF) power amplifiers operating near saturation and from low-resolution data converters, significantly impacting performance in systems such as massive multiple-input multiple-output (MIMO), millimeter-wave (mmWave), and integrated sensing and communications (ISAC). With increasing used of mmWave frequencies for vehicular communication systems, narrow-beam transmission and accurate beam tracking are essential for maintaining reliable connectivity under high mobility. However, prior studies have shown that power-amplifier nonlinearities can substantially degrade MIMO beamforming performance by introducing amplitude and phase distortions that alter the intended beamforming strategy and beam patterns \cite{Qi_TCOM_2012,Shaham_IEEEAccess_2019,Shaham_CL_2020}. Consequently, accurately characterizing the statistical effects of hardware nonlinearities is critical for the design of robust next-generation wireless systems.

In this paper, we model the output of a nonlinear RF power amplifier circuit as a function of its input signal and noise, where the desired signal is applied at the RF input and the bias circuitry introduces a baseband noise component \cite{kobal,kumar,kobal2}. Treating both the signal and noise as stochastic processes, we aim to derive a numerical characterization of the output spectrum in the vicinity of each harmonic of the input.
Since the proposed characteristic-function (ch.~f.) framework quantifies the statistical properties of nonlinear distortion and noise propagation through RF front ends, it provides a complete statistical description of the received signal in the presence of hardware impairments. Such information can be exploited to develop distortion-aware communication algorithms, including likelihood-based data detection, maximum a posteriori (MAP) receivers, and channel estimation techniques that account for non-Gaussian distortion statistics rather than relying on conventional Gaussian assumptions \cite{li2017,bjornson2018,ding2025}. Furthermore, the framework enables the evaluation of key performance metrics such as signal-to-distortion ratio (SDR), bit error rate (BER), and outage probability without extensive Monte Carlo simulations. Beyond receiver design, the proposed methodology can support distortion-aware beam management, beam tracking, and reliability analysis in future vehicle-to-everything (V2X), 6G, and other beamformed wireless communication systems affected by nonlinear RF front-end impairments \cite{Aghdam_TSP_2021}. 

The problem at hand can be abstracted to that of nonlinear operations on stochastic processes, a subject which is condensed in the literature in a few places, Rice being a primary source \cite{rice1,rice2}. Three chapters of \cite{blachman} are devoted to the treatment of stochastic processes in nonlinear systems, especially expanding upon the Chebyshev transform. In \cite[chapter 9]{cattermole}, the authors address such analysis using power series, orthogonal expansions, and ch. f., the latter being noted for its generality and therefore wide applicability. In some earlier works \cite{jain,abuelmaatti1,abuelmaatti2,zhang}, several tools are used to tackle similar problems, including the use of Fourier series to represent a nonlinearity. In \cite{demir2020, dinis,rajaram}, researchers applied Bussgang's theorem \cite{bussgang} (a special case of Price's theorem \cite{price}) to attack the problem, though this technique is limited to Gaussian processes. The novelty of our work lies in reformulating Rice's ch. f., method using Fourier series as the representation of the nonlinearity, resulting in a more general approach to this subject that is more amenable to numerical evaluation, and applicable to stochastic processes of any distribution since the ch. f., of a wide-sense stationary random voltage or current always exists. It should be noted that the proposed framework is not intended to replace Volterra-series methods; rather, it addresses a different class of problems by providing a tractable ch.~f., analysis for memoryless nonlinear systems with stochastic inputs.

The rest of this paper is organized as follows. Section II presents the problem statement and assumptions. Section III presents a novel derivation of the output correlation with an objective of facilitating numerical computation of the output terms, and Section IV summarizes the steps involved in computing it. Section V presents a numerical example and lastly, Section IV concludes the paper.
\section{Problem Statement}
Consider two signals, applied at the input of a nonlinear system, $x_1(t)$ and $x_2(t)$ consisting of a sinusoid-plus-noise as
\begin{subequations}\label{deqn_e2}
\begin{align}
x_1(t) &=P \cos(\omega t+\theta)+n_1(t)+\beta,\\
x_2(t) &=P\cos(\omega t + \theta)+n_2(t)+\beta,
\end{align}
\end{subequations}
where $P$ and $\omega$ are constants; $n_1(t)$ and $ n_2(t)$ are jointly Gaussian processes with zero mean, variances $\sigma_1^2, \sigma_2^2$, and correlation coefficient $\rho(\tau)=\mathbb{E}\{n_1(t+\tau)n_2(t)\}/\sigma_1 \sigma_2$, with time lag $\tau$; $\theta$ is a uniform random variable in $(-\pi,\pi)$, and independent of $n_1, n_2$ and $\beta$ is a \texttt{dc} bias on the input signals. $\mathbb{E}\{\cdot\}$ represents the expectation operator. The input signals then pass through memoryless nonlinear distortions represented by the functions $f(x)$ and $g(x)$, and result in the following output signals:
\begin{equation}\label{deqn_e1}
y_1(t) =f\left(x_1(t)\right),\quad
y_2(t) =g\left(x_2(t)\right).
\end{equation}
Accordingly, the input signal-to-noise ratio (SNR) can be deduced as $P^2/(2\sigma_1\sigma_2)$. The uniform random phase of the signal, $\theta$, embodies the requirement that the signal itself is random, but ergodic, so that signal statistics are independent of parameter $t$. Moreover, it will become evident later that the requirement for Gaussian noise is convenient, but that the analysis can be used with other noise distributions as desired.

The primary input variable is the input noise spectrum by means of its transform, the correlation coefficient $\rho(\tau)$, and the problem at hand is to obtain the spectrum of the output by calculating the cross correlation $\Psi(\tau)$ between samples of the output signals $y_1(t)$ and $y_2(t)$. We will begin with the derivation for a general two-input, two-output system, and then make it particular for a single input, single output system wherein we are given the input autocorrelation, and seek the output autocorrelation. 




\section{Derivation of Cross Correlation $\Psi(\tau)$}
This section describes the step-by-step procedure to obtain the cross correlation function $\Psi(\tau)$. 
To this end, we write the expression for output correlation $\Psi(\tau)$ as the expected value of the product of the outputs that incorporates the nonlinearities (\ref{deqn_e1}) and input signals (\ref{deqn_e2}) as
\begin{eqnarray}\label{deqn_e3}
\Psi(\tau)=\mathbb{E}\left\{ f \left( P \cos{\left(\omega (t+\tau)+\theta \right)} + n_1(t+\tau) +\beta\right) \right. \notag\\
\times \left. g\left(P \cos{(\omega t + \theta)} + n_2(t) + \beta\right)\right\}.
\end{eqnarray}
To proceed further, we need to express the nonlinear transfer functions $f(x)$ and $g(x)$ in a tractable form. A power series is an option \cite{cattermole}, while \cite[section 4.8]{rice2} used Fourier transforms, which ultimately require computation of multiple nested improper integrals. These approaches are general enough to accommodate inputs of infinite amplitude, but in practice, the input signal amplitude is finite, so it is entirely valid to define the nonlinearity in a more convenient form as long as it corresponds to the actual nonlinearity over the range of the input signal. This observation suggests that a more tractable form may be obtained by employing a Fourier series \cite{churchill} of a periodic extension of the nonlinearity instead of the Fourier transform of the actual nonlinearity. This results in an expression that is more easily computed than the improper integrals. We hasten to point out that the nonlinearity need not \emph{actually} be periodic, such as an Mach-Zehnder
modulator (MZM) \cite{mzm}, merely that the periodic extension corresponds with the actual nonlinearity over the range of the input variable of interest. Strictly speaking, a Gaussian r.v. is unbounded, but if the analyst is satisfied with ignoring the effects of large excursions occurring with infinitessimal probability, then the practicality argument is persuasive. 

We assume that $f$ and $g$ are periodic with period $2c$, and can be expressed as a Fourier series (\ref{deqn_e4}). (Strictly speaking, $f$ and $g$ are defined in the limit as the sum approaches an infinite number of terms, but we omit the limit nomenclature for brevity. The limit operation is implicit in (\ref{deqn_e4}) and following.)
\begin{subequations}\label{deqn_e4}
\begin{align}
f(x) =  \sum_{q=-\infty}^{\infty} F_q e^{j \frac{q \pi x}{c}} , \quad F_q = \frac{1}{2c} \int_{-c}^{c} f(x) e^{-j\frac{q \pi x}{c}} dx,\\
g(x) = \sum_{r=-\infty}^{\infty} G_r e^{j \frac{r \pi x}{c}},\quad G_r = \frac{1}{2c} \int_{-c}^{c} g(x) e^{-j\frac{r \pi x}{c}} dx, 
\end{align}
\end{subequations}
where $F_q$ and $G_r$ are Fourier series coefficients. It is worth emphasizing that these series are merely a mathematical device used to facilitate the computation of the expectation operator in \eqref{deqn_e3} and have no relation to the frequency response of the nonlinear device, nor to the spectral content of the signals or outputs. 

Next, substituting (\ref{deqn_e4}) into (\ref{deqn_e3}), we can write 
\begin{multline}
\label{deqn_e5}
\Psi = \mathbb{E} \left\{ \left( \sum_{q=-\infty}^{\infty} F_q e^{j \frac{q\pi}{c} [P \cos{(\omega(t+\tau)+\theta)+n_1(t+\tau)+\beta]}}\right) \right. \\
\times \left. \left( \sum_{r=-\infty}^{\infty} G_r e^{j \frac{r\pi}{c} [P \cos{(\omega t+\theta)+n_2(t)+\beta]}}\right) \right\}.
\end{multline}
We can rearrange \eqref{deqn_e5} by leveraging the independence of $\theta$ and $n_1, n_2$ as
\begin{multline}
\label{deqn_e6}
\Psi =  \sum_{q=-\infty}^{\infty} \sum_{r=-\infty}^{\infty} \underbrace{F_q G_r}_\text{nonlinearity} \quad \underbrace{e^{j(\frac{q\pi}{c} + \frac{r\pi}{c})\beta}}_\text{ch. f. of bias, $\Phi_\beta(q,r)$}\\
\times \underbrace{\mathbb{E}_\theta \left\{ e^{j \frac{q\pi}{c} [P \cos{(\omega(t+\tau)+\theta)]}} e^{j \frac{r\pi}{c} [P \cos{(\omega t+\theta)]}} \right\} }_\text{ch. f. of sinusoid, $\Phi_C(q,r)$}\\
\times \underbrace{\mathbb{E}_{n_1,n_2}\left\{ e^{j \frac{q\pi}{c} n_1(t+\tau)} e^{j \frac{r\pi}{c} n_2(t)} \right\}}_\text{ch. f. of noise}.
\end{multline}
Since the expected value of a nonrandom variable is just that variable itself, the Fourier coefficients and the exponential function of the bias voltage are unchanged, leaving just the expectation operations on the random-phase sinusoid and the noise. Recalling that $\mathbb{E}_X\{e^{jX}\}$ is the ch. f., of r.v. $X$ \cite{peebles}, we observe that the first expectation is the joint ch. f., for the random variables $P\cos{(\omega (t+\tau) + \theta)}$ and $P\cos{(\omega t +\theta)}$, which we will denote $\Phi_{C} (q,r )$. The second expectation is the joint ch. f., of $n_1(t+\tau)$ and $n_2 (t)$, and thus, we observe that (\ref{deqn_e6}) is similar to Rice’s fundamental formula of the ch. f., method \cite[equation 4.8-6]{rice2}. 

Next we generalize \eqref{deqn_e6} using the Fourier series coefficients of the nonlinearities and sampling the general ch. f. of the signals and noise, thereby introducing the \emph{series form of the fundamental formula of the ch. f. method}. For nonlinearities with periods $2c$ and $2d$, described with Fourier series coefficients $F_q$ and $G_r$, respectively, we write
\begin{equation}
\label{deqn_e7}
\Psi(\tau) =  \sum_{q=-\infty}^{\infty} \sum_{r=-\infty}^{\infty} F_q G_r \phi \left( \frac{q\pi}{c}, \frac{r\pi}{d}, \tau \right),
\end{equation}
where $\phi(u,v,\tau)$, with $u=q\pi/c$ and $v=r\pi/d$, denotes the generalized ch. f., of the input process, which is the product of ch. fs., of all the input terms (i.e., the \texttt{dc} bias, the sinusoid and the noise). Additional input terms would be handled by including each ch. f. as a factor in $\phi(u,v,\tau)$. In our case,  $\phi(u,v,\tau)=\Phi_\beta(\frac{q\pi}{c},\frac{r\pi}{c},\tau)\,\Phi_{C}(\frac{q\pi}{c},\frac{r\pi}{c},\tau) \, \Phi_{n_1 n_2}(\frac{q\pi}{c},\frac{r\pi}{c},\tau)$. (We will henceforth omit $\tau$ from the argument lists for brevity sake.) We proceed by explicitly writing the ch. f. of the noise terms because they are jointly Gaussian random variables \cite[page 336]{peebles} as
\begin{align}
\begin{split}
\label{deqn_e8}
\mathbb{E}_{n_1,n_2} &\left\{  e^{j \frac{q\pi}{c} n_1(t+\tau)} e^{j \frac{r\pi}{c} n_2(t)} \right\} \\
&= e^{-\frac{\pi^2}{2c^2} [\sigma_1^2 q^2 + 2\sigma_1 \sigma_2 qr\rho + \sigma_2^2 r^2]} \\
&= e^{-\frac{\pi^2}{2c^2} (\sigma_1^2q^2+\sigma_2^2r^2)} \sum_{k=0}^{\infty} \frac{(-1)^k \left(\frac{\pi}{c}\right)^{2k} (\sigma_1 \sigma_2 qr\rho)^k}{k!},
\end{split}
\end{align}
where the exponential function in \eqref{deqn_e8} is expanded as its Maclaurin series polynomial in $\rho(\tau)$. For the case where the noise process is actually a different distribution, we need to write the joint ch. f., of that process instead. Substituting \eqref{deqn_e8} into \eqref{deqn_e6}, and rearranging the order of summation, we rewrite $\Psi$ as
\begin{multline}
\label{deqn_e9}
\Psi = \sum_{k=0}^{\infty} \underbrace{\frac{(-1)^r \left( \frac{\pi}{c} \right)^{2k} (\sigma_1 \sigma_2 \rho)^k}{k!}}_\text{noise} \\
\times  \sum_{q=-\infty}^{\infty} \sum_{r=-\infty}^{\infty} \underbrace{F_q G_r}_\text{nonlinearity} \underbrace{e^{j(\frac{q\pi}{c} + \frac{r\pi}{c})\beta}}_\text{bias}\\
\times \underbrace{\Phi_C(q,r)}_\text{sinusoid} \underbrace{e^{-\frac{\pi^2}{2c^2} (\sigma_1^2q^2+\sigma_2^2r^2)} (qr)^k}_\text{noise}.
\end{multline}

We now turn our attention to finding an expression for $\Phi_{C}(q,r)$.  Since $\theta$ is a uniformly distributed r.v. over $(-\pi,\pi)$, we can write
\begin{align}
\begin{split}
\label{deqn_e10}
\Phi_C(&q,r) = \mathbb{E}_\theta \left\{ e^{j\frac{q\pi}{c}[P\cos{(\omega(t+\tau)+\theta)]}} e^{j\frac{r\pi}{c}[P\cos{(\omega t+\theta)]}} \right\}, \\
&=\frac{1}{2\pi} \int_{-\pi}^{\pi} e^{j\frac{q\pi}{c}[P\cos{(\omega(t+\tau)+\theta)]}} e^{j\frac{r\pi}{c}[P\cos{(\omega t+\theta)]}} \,d\theta .
\end{split}
\end{align}
 By the change of variable from $(\omega t+\theta)$ to $\alpha$, it becomes obvious that $\Phi_{C}(q,r)$ is independent of $t$. If we let $\frac{q\pi}{c} P = a$ and $\frac{r\pi}{c} P = b$, it can easily be shown that
\begin{equation}\label{deqn_e11}
a\,\cos{(\omega\tau+\alpha)}+b\,\cos{(\alpha)}=\sqrt{a^2+b^2+2ab\cos{(\omega\tau)}} \cos{(\alpha+\eta)}.
\end{equation}

Parameter $\eta$ can be computed from the arctangent of the factors of terms involving $\sin{\alpha}$ and $\cos{\alpha}$, but is unimportant because of the periodicity of the integral. Substituting \eqref{deqn_e11} into \eqref{deqn_e10}, we observe that the integral is the definition of the zeroth order Bessel function of the first kind: $J_0(z)=(2\pi)^{-1}\int_{-\pi}^{\pi} \exp (j z \sin u) \, du$ \cite[page 186]{churchill}. Accordingly, $\Phi_{C}(q,r)$ can be expressed as
\begin{align}
\begin{split}
\label{deqn_e12}
\Phi_{C}(q,r)
&=\frac{1}{2\pi}\int_{-\pi}^\pi e^{j \sqrt{a^2+b^2+2ab\cos{(\omega\tau)}} \cos{(\alpha+\eta)}} d\alpha,\\
&=\frac{1}{2\pi}\int_{-\pi}^\pi e^{j \sqrt{a^2+b^2+2ab\cos{(\omega\tau)}} \sin{(\beta)}} d\beta,\\
&=J_0 \left( \frac{\pi P}{c} \sqrt{q^2 + 2qr \cos{(\omega \tau)} + r^2} \right). \\
\end{split}
\end{align}
We further employ a useful variation on Neumann’s Addition Formula for Bessel functions \cite[page 358]{watson1922treatise} to expand \eqref{deqn_e12} into a series, such that \eqref{deqn_e12} can be rewritten as the real quantity, 
\begin{multline}
\label{deqn_e13}
\Phi_C(q,r)=\sum_{n=0}^{\infty} \epsilon_n J_n \left(\frac{q\pi}{c} P \right) J_n \left(\frac{r\pi}{c} P \right) (-1)^n \cos{(n \omega \tau)},
\end{multline}
where 
$
\epsilon_n = 
\begin{cases}
1, &n=0 \\
2, &n\neq0
\end{cases}
$, and $J_n$ is the Bessel function of $n$th order. The ch. f. of the sinusoid is now expressed as a sum of harmonics of the input frequency, weighted by Bessel functions. This will become very useful as we proceed. We now substitute \eqref{deqn_e13} into \eqref{deqn_e9}, and rearrange into separate sums over $q$ and $r$ as
\begin{multline}
\label{deqn_e14}
\Psi = \sum_{n=0}^{\infty} \sum_{k=0}^{\infty}
\left[  \sum_{q=-\infty}^{\infty} F_q  e^{j\frac{q\pi\beta}{c}} J_n \left( \frac{q\pi}{c} P\right) e^{-\frac{\pi^2}{2c^2} \sigma_1^2 q^2} \left( \frac{q\pi}{c} \right)^k \right] \\
\times \left[  \sum_{r=-\infty}^{\infty} G_r e^{j\frac{r\pi\beta}{c}} J_n \left( \frac{r\pi}{c} P\right) e^{-\frac{\pi^2}{2c^2} \sigma_2^2 r^2} \left( \frac{r\pi}{c} \right)^k \right] \\
\times \frac{\epsilon_n(-1)^n (-1)^k}{k!} (\sigma_1 \sigma_2 \rho)^k \underbrace{\cos{(n \omega \tau)}}_\text{harmonics of $\omega$}.
\end{multline}
It is observed from expression \eqref{deqn_e14} that we have successfully decomposed the correlation function $\Psi$ into a harmonic series of the fundamental frequency $\omega$ of the original sinusoid. The series is indexed by harmonic number $n$ and noise exponent $k$. 

It is straightforward to {apply this result to the single input, single output case $f(x)=g(x)$, $\sigma_1=\sigma_2=\sigma$, and $\rho(\tau) $ as the autocorrelation coefficient of the input noise process. We then write \eqref{deqn_e14} as an series expansion where a polynomial in $\rho$ scales each harmonic function $\cos{n \omega \tau}$
\begin{equation}
\label{deqn_e16}
\Psi= \sum_{n=0}^{\infty} \sum_{k=0}^{\infty} \epsilon_n h_{n,k}^2 \frac{(\sigma^2 \rho)^k}{k!} \cos{(n \omega \tau)},
\end{equation}
where the constant coefficients $h_{n,k}^2$ are parameterized by signal amplitude $P$, \texttt{dc} bias $\beta$, and noise variance $\sigma^2$ as
\begin{multline}
\label{deqn_e17}
h_{n,k}=
(j)^{n+k}  \sum_{\lambda=-\infty}^{\infty} F_\lambda \underbrace{e^{j\frac{\lambda\pi\beta}{c}}}_\text{bias} \underbrace{J_n \left( \frac{\lambda \pi P}{c} \right)}_\text{sinusoid} 
\underbrace{e^{-\frac{\pi^2}{2c^2}\sigma^2 \lambda^2} \left( \frac{\lambda \pi}{c} \right)^k}_\text{noise}\\
n,k \ge 0
\end{multline}

A remark is in order regarding the $\rho^k=\rho^k(\tau)$ term in \eqref{deqn_e16}. Since the spectrum is the Fourier transform of the correlation, then the convolution theorem of Fourier transforms tells us that the spectrum of $\rho^{k}$ is the $k$th autoconvolution of the input spectrum, as shown in Fig. \ref{fig5} for an ideal bandlimited noise. This produces an output spectrum consisting of an impulse weighted by $h_{1,0}^2$ at the frequency of the input sinusoid $\omega$ (because $\rho^0=1$), which is the output signal, superposed with the noise spectrum convolved with itself $(k-1)$ times weighted by $h_{1,k}^2 \sigma^{2k} /k!$ and centered at $\omega$. If the bandwidth of the noise is considerably smaller than the frequency of the sinusoid, then a good approximation of the spectrum at each harmonic frequency can be obtained in isolation from the others. 
\begin{figure}[!t]
\centering
\includegraphics[width=3.5in]{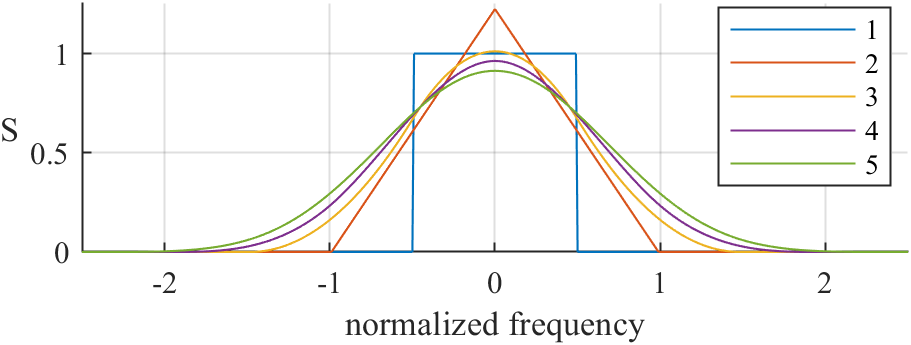}
\caption{Spectra of the noise contributions in \eqref{deqn_e16} for $k=1,2,...5$, based on the ideal bandlimited input spectrum of a normalized bandwidth.}
\label{fig5}
\end{figure}

\section{Summary of the Procedure}
It is important to emphasize that the foregoing derivation can be applied to a wide variety of nonlinearities (including those that arise due to power amplifiers, low-resolution analog-to-digital converter and finite-resolution quantization) and signals-plus-noise. This generic procedure is summarized as follows:
\begin{enumerate}
\item Devise a suitable $2c$-periodic extension of the nonlinear transfer function for the anticipated input signal conditions, particularly the peak magnitude of signals-plus-noise. Compute the Fourier series coefficients $F_\lambda$.
\item{Compute the general ch. f. of the noise, as in (\ref{deqn_e8}), which in many circumstances may be the continuous ch. f. $\phi(u,v,\tau)$ evaluated at discrete values $u=q\pi/c$ and $v=r\pi/c$. It will be particularly helpful to expand the ch. f. as a polynomial in $\rho(\tau)$ to facilitate relating the output spectrum to the input spectrum.}
\item{Compute the general ch. f. of the sinusoid, which in many cases may be \eqref{deqn_e13}. Multiple sinusoids may also be included, each with its own ch. f.}
\item{Combine all the ch. fs. with the Fourier coefficients as in (\ref{deqn_e14}), and simplify to arrive at an expression for the harmonic outputs and weighting coefficients $h_{n,k}$ as in \eqref{deqn_e16} and \eqref{deqn_e17}.}
\item{Evaluate the weights $h_{n,k}$, and determine which values of $k$ contribute significantly to the output. Determine the spectral shapes of the $k^{th}$ autoconvolutions of the input (e.g. Fig. \ref{fig5}, etc.). Sum and weight these spectra to arrive at the net output spectrum at the selected harmonic.}
\end{enumerate}

\section{Numerical Results}
\noindent Consider the transconductance of a GaN HEMT shown in Fig. \ref{fig1} adapted from \cite{lee}. We desire to analyze the behavior of this device in class-AB mode, with the signal biased near the peak of the G\textsubscript{m} curve at $\text{V}_\text{GS}=-2.5\text{V}$. We hypothesize that the system involves a sinusoid input, and a baseband noise term that is coupled in to the bias network. Equation \eqref{deqn_e16} tells us that the output will consist of a sinusoid spectrum (delta function) superposed with a noise spectrum. We are interested in the spectrum around the fundamental frequency of the input sinusoid, so $n=1$. We compute the Fourier series for the periodic extension of the nonlinearity shown in Fig. \ref{fig2} which is exactly equal to the actual nonlinearity over the interval $-11<\text{V}_\text{GS}<0$. We choose $2c=14\text{V}$ for the period. If analysis for a larger input is required, it is a simple matter to draw a different periodic extension, producing a different set of constant coefficients $F_\lambda$.

\begin{figure}[!t]
\centering
\includegraphics[width=3.5in]{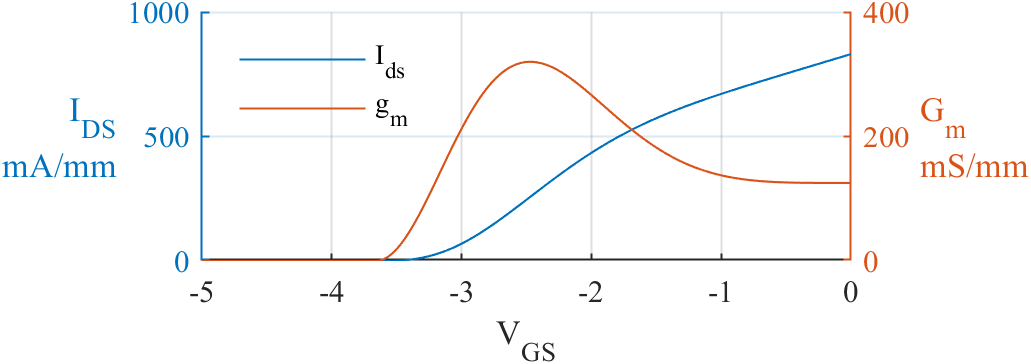}
\caption{Nonlinear transfer function of GaN HEMT, adapted from \cite{lee}.}
\label{fig1}
\end{figure}

\begin{table}
\begin{center}
\caption{I\textsubscript{ds} polynomial coefficients.}
\label{tab1}
\begin{tabular}{| c | r | c |}
\hline
Coefficient & Value & \\
\hline
 $a_0$ & 837.1 & \multirow{4}{*}{ $ \displaystyle \text{I}_\text{DS} \approx\sum_{n=0}^5 a_n (\text{V}_\text{GS})^n, $} \\
 $a_1$ & 150.8  &\\
 $a_2$  & -40.44  &\\
 $a_3$  & -65.55 &\\
 $a_{4}$  & -45.00 & \multirow{2}{*}{ $ \displaystyle -3.5 < \text{V}_\text{GS} < 0$}\\
 $a_{5}$  & -7.870 &\\
\hline
\end{tabular}
\end{center}
\end{table}

Since we are free to express $f(x)$ in any way that is convenient for the computation of \eqref{deqn_e4}, we chose a polynomial with the coefficients in Table \ref{tab1} defined only for $-3.5 < \text{V}_\text{GS} < 0$, and zero otherwise. Computation of the Fourier coefficients of the periodic function in Fig. \ref{fig2} where the nonzero portions are described by the polynomial is a straightforward task. 

\begin{figure}[!t]
\centering
\includegraphics[width=3.5in]{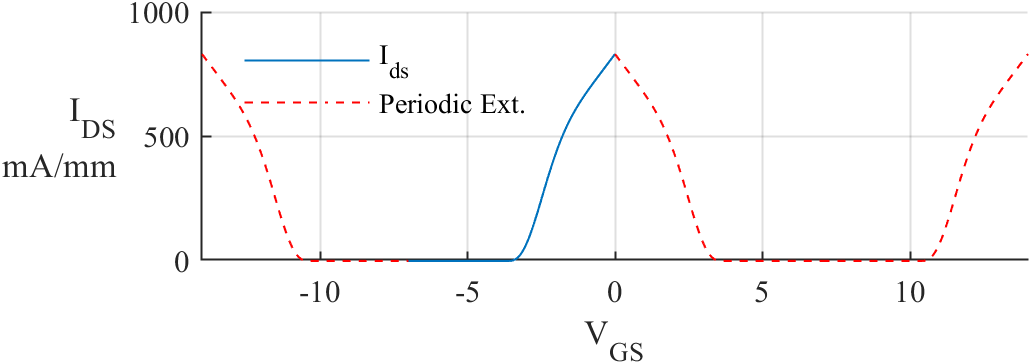}
\caption{Periodic extension of the I\textsubscript{DS} vs. V\textsubscript{GS} curve in Fig. \ref{fig1}, period $2c=14\text{ Volts}$.}
\label{fig2}
\end{figure}

This procedure provides a powerful quantitative result. For purposes of this work, however, we make a few qualitative observations. We fix SNR=40 dB, and perform the calculations. When the component powers $\sigma^{2k}h_{1,k}^2/k!$ are calculated across a range of sinusoid input amplitude $P$ (Fig. \ref{fig3}), we see that the sinusoid $k=0$ term begins to compress as expected for large $P$, and that the noise terms $k>0$ rapidly diminish in significance for higher order. Looking a little closer at the most significant noise term ($k=1$), we can see the effect of the bias voltage on the overall noise power (Fig. \ref{fig4}). As bias voltage deviates from the peak of the gain G\textsubscript{m} at -2.5V, the noise power increases considerably, as well as changing shape where the noise experiences localized nulls. This expansion provides a tool to explore this phenomenon more thoroughly in further research.

\begin{figure}[!t]
\centering
\includegraphics[width=3.5in]{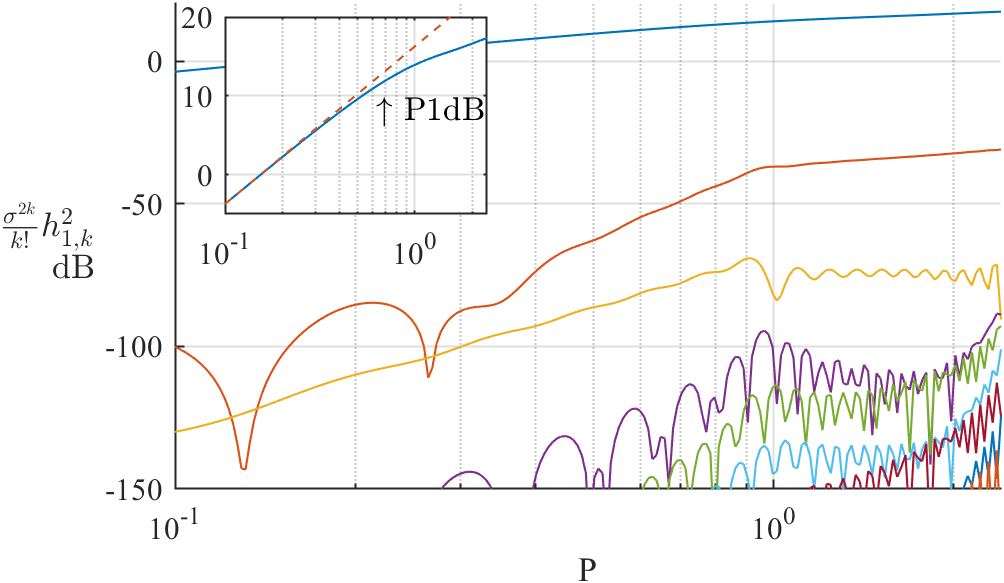}
\caption{Expansion coefficients vs. input sinusoid amplitude $P$, calculated as $\sigma^{2k}h_{1,k}^2/k!$ \eqref{deqn_e17} for $k=0,1,...,8$, which appear in order from top to bottom, for SNR=40 dB. Inset shows $h_{1,0}^2$ and 1-dB compression point.}
\label{fig3}
\end{figure}

\begin{figure}[!t]
\centering
\includegraphics[width=3.5in]{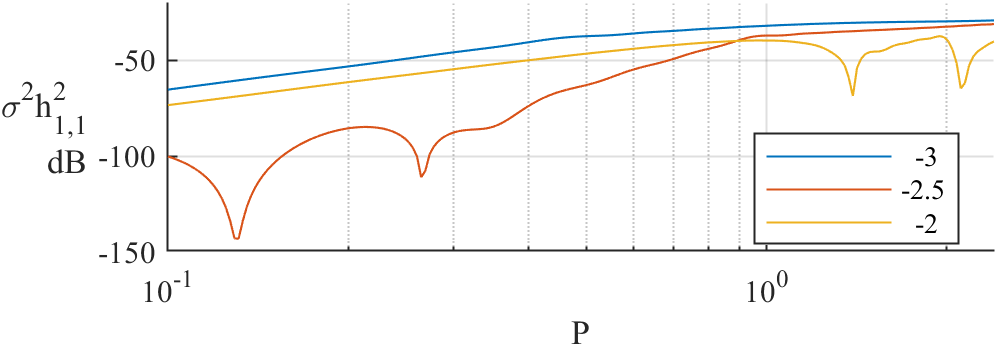}
\caption{Expansion coefficient $\sigma^2h_{1,1}^2$ \eqref{deqn_e17} for $V_{bias}=\beta=-3.0, -2.5, -2.0$V, SNR=40 dB.}
\label{fig4}
\end{figure}

\section{Conclusion}
This paper addressed the problem of statistically characterizing signal and noise propagation through memoryless nonlinear RF devices, which has become increasingly important in modern wireless systems where hardware impairments can affect beamforming accuracy, beam tracking performance, signal-to-distortion ratio, and link reliability. To this end, we extended Rice’s ch.~f., method by representing the nonlinearity through a Fourier-series expansion of its periodic extension, thereby transforming computationally intensive multidimensional improper integrals into tractable summations. The resulting framework enabled efficient characterization of nonlinear distortion and noise propagation while retaining applicability to arbitrary signal and noise distributions. Numerical results based on a GaN HEMT demonstrated the ability of the proposed method to capture the interplay among signal amplitude, bias voltage, and noise, including operating regions where nonlinear effects partially suppressed noise contributions. Beyond the illustrative device example, the proposed framework provided a general analytical tool for studying hardware impairments in massive MIMO, mmWave, ISAC, vehicular, and future 6G communication systems. Future work will focus on extending the framework to modulated signals, experimentally validating the analytical predictions, and investigating non-Gaussian signal and noise distributions for distortion-aware wireless system design.

\bibliographystyle{IEEEtran}
\bibliography{nonlinear_fourier}

\end{document}